\documentclass{iaus}
\usepackage{graphics}

  \checkfont{eurm10}
  \iffontfound
    \IfFileExists{upmath.sty}
      {\typeout{^^JFound AMS Euler Roman fonts on the system,
                   using the 'upmath' package.^^J}%
       \usepackage{upmath}}
      {\typeout{^^JFound AMS Euler Roman fonts on the system, but you
                   dont seem to have the}%
       \typeout{'upmath' package installed. iaus.cls can take advantage
                 of these fonts,^^Jif you use 'upmath' package.^^J}%
      }
  \else
  \fi

  \checkfont{msam10}
  \iffontfound
    \IfFileExists{amssymb.sty}
      {\typeout{^^JFound AMS Symbol fonts on the system, using the
                'amssymb' package.^^J}%
       \usepackage{amssymb}%
         \let\leq=\leqslant
         
      }{}
  \fi

  \IfFileExists{amsbsy.sty}
    {\typeout{^^JFound the 'amsbsy' package on the system, using it.^^J}%
     \usepackage{amsbsy}}
    {\providecommand\boldsymbol[1]{\mbox{\boldmath $##1$}}}

\newsavebox{\astrutbox}
\sbox{\astrutbox}{\rule[-5pt]{0pt}{20pt}}

\title[Newcomers Meet the Intracluster Medium in Coma]
      {Newcomers Meet the Intracluster Medium in the Coma Cluster}

\author[B. M. Poggianti {\it et al.\/}]
{Bianca M. Poggianti$^1$, T.J.\ Bridges$^2$, M. Yagi$^3$, Y. Komiyama$^4$, D. Carter$^5$, B. Mobasher$^6$, S. Okamura$^7$, N. Kashikawa$^3$}
\affiliation{$^1$ INAF-Osservatorio Astronomico di Padova, Italy  \\[\affilskip] $^2$ Anglo-Australian Observatory, Australia \\[\affilskip]  $^3$ National Astronomical Observatory, Mitaka, Tokyo, Japan \\[\affilskip] $^4$ Subaru Telescope, Hilo, HI, USA \\[\affilskip] $^5$ Liverpool John Moores University, Birkenhead, Wirral, UK \\[\affilskip] $^6$ Space Telescope Science Institute, Baltimore, USA \\[\affilskip] $^7$ Department of Astronomy, University of Tokyo, Japan }
\pubyear{2004}
\volume{195}
\pagerange{1--8}
\date{?? and in revised form ??}
\jname{Outskirts of Galaxy Clusters: intense life in the suburbs}
\editors{A. Diaferio, ed.}
\begin{document}
\maketitle
\begin{abstract}
A main topic at this meeting is 
how galaxies are affected when they {\it enter} for the first time
the cluster environment from the outskirts. Most of the times we are forced
to infer the environmental effects indirectly, relying on systematic
variations of galaxy properties with environment, 
but there aren't many examples of direct observations 
able to unveil ongoing transformations taking place, and the
corresponding mechanism producing it.
We present a case in which it is
possible to identify the cluster environment, and in particular the
intracluster medium and the recent infall history of galaxies onto the
cluster, as the cause for a recent, abrupt change in the evolutionary history
of galaxies. 
\end{abstract}

\section{Post-starburst galaxies and substructure in Coma}

This study is based on a photometric and 
spectroscopic survey of galaxies in the Coma cluster
(Mobasher et al. 2001), which is distinctive from other surveys in three 
ways: for the large galaxy magnitude range covered (almost 7 mag, down
to $M_B \sim -14$), for the large area surveyed (two regions, towards the 
cluster center and to the South-West, $\sim 1 \times 1.5$ Mpc each),
 and for being a simply magnitude limited sample, with
no morphological selection criteria adopted for spectroscopy.
In this survey
we have found that a significant fraction ($\sim 10$ \%) of the
cluster dwarf galaxy population at $M_V>-18.5$ 
has post-starburst/post-starforming spectra (Poggianti et al. 2004, hereafter
P04).  
This type of spectrum (``k+a'', or ``E+A'') indicates a
galaxy with no current star formation activity which was forming stars
at a vigorous rate in the recent past (last 1.5 Gyr).
In the $B-R$ color-magnitude diagram, a group of blue and a group of red
k+a's can be easily distinguished in Coma.  The average EW($\rm
H\delta$) of the blue group is significantly stronger than that of the
red group.  The blue, strong k+a's most likely correspond to ``young''
k+a's (observed soon after the termination of star formation, $<$ 
300 Myr) and the
red, weaker k+a's are ``old'' ones (observed at a later stage of the
evolution, 0.5-1.5 Gyr).

\begin{figure}
\includegraphics{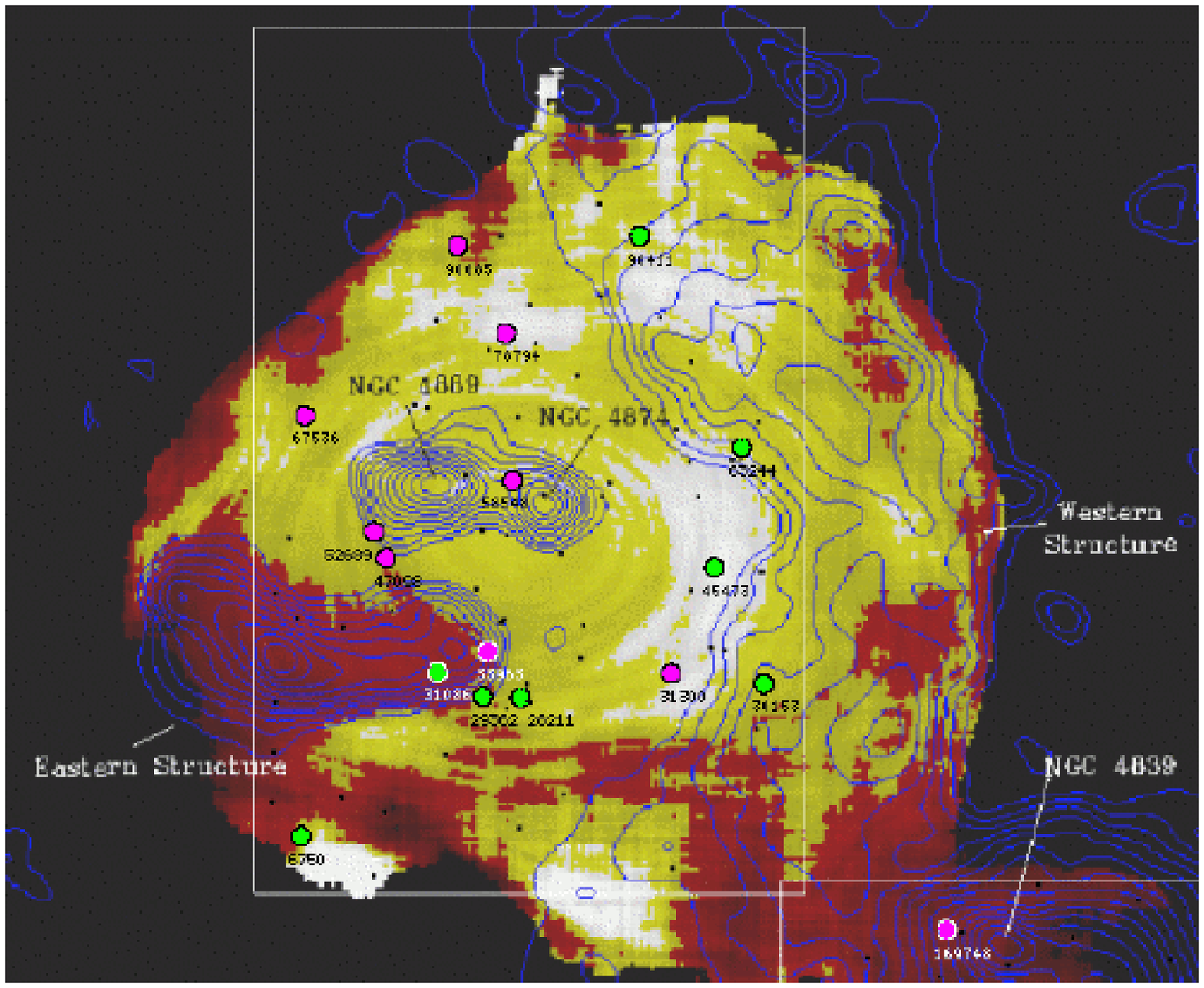}
\caption{{\it N.B. This plot is best viewed in color, see Poggianti et al. 
2004.} Position of k+a galaxies with respect to X-ray substructure
and X-ray temperature map. Only the central field of Coma is shown
here, see P04 for a full map. Strong-lined k+a's with EW($\rm
H\delta)>5$ \AA $\,$ are shown as light-coloured large dots, 
while weaker k+a's are
plotted as darker large dots.  Tiny black dots are dwarf Coma members with
velocities $> 7200 \, \rm km \, s^{-1}$.  X-ray residuals from
Neumann et al. (2003) are plotted as contours and clearly identify two
substructures (Western and Eastern substructures), in addition to the
NGC4839 peak in the South-West and the excess of emission towards the
two central galaxies (NGC4874 and NGC4889).  The lowest contour and
the step width between two contours are each 5 $\sigma$.  The hardness
ratio image (2-5 keV/0.5-2keV, Neumann et al. 2003) is also shown. Darker
regions correspond to temperatures below 8 keV, intermediate-grey regions 
to $kT>8$ keV
and light-coloured regions to $kT>10$ keV.  The rectangles show the limits of
the two fields of our photometric and spectroscopic survey (Coma1
towards the cluster center and Coma3 in the South-West). The
rectangle is about 1 by 1.5 Mpc.}
\end{figure}

A suggestive clue about the possible physical mechanism responsible
for the k+a spectra comes from the recent X-ray mosaic observations of
Coma obtained with {\it XMM-Newton}. Coma has two central dominant
galaxies, NGC 4874 (a cD galaxy) and NGC4889 (a very bright elliptical), 
and another cD galaxy, NGC4839,
that dominates a substructure South-West of the center (Fig.~1).  Neumann
et al. (2003) have identified and discussed X-ray
substructure by fitting a smooth profile and subtracting it from the
data. The residuals reveal several structures, that are shown as
contours in Fig.~1: besides the well known NGC4839 South-West group,
Neumann et al. find a large residual to the West of the cluster
centre (``Western structure'' in Fig.~1) elongated along the
North-South direction, and a filament-like structure South-East of the
centre (``Eastern structure'' in Fig.~1), elongated along the
East-West direction.  The temperature map shown in Fig.~1
sheds further light on the accretion history of Coma. Neumann et
al. conclude that the region of high temperature observed between the
Western structure and the Coma center is caused by the infall of this
structure, either via compression or via shock waves.  These authors
consider the two maxima in the western structure to be likely the
result of the disruption of a galaxy group during its infall, instead
of two galaxy groups falling at the same time.  In contrast, the
South-Eastern structure is cooler than the mean cluster temperature
and is associated with a low-mass galaxy group dominated by two large
galaxies, NGC4911 and NGC4921.  Based on the filamentary form of this
structure, the same authors conclude it is observed during the infall
process while being affected by ram pressure stripping close to the
cluster centre.

The coincidence of the position of the strongest k+a galaxies and the
X-ray structures is striking. Four k+a's with EW($\rm H\delta)>5$ \AA
$\,$ (light large dots in Fig.~1) trace the edge of the Western structure
towards the Coma centre. Another three are associated with the
Eastern structure, all at its western boundary.  
Thus, young post-starbursts are distributed
close to the edge of infalling substructures. In the case of the Western
substructure this edge is the infalling front, while for the Eastern
substructure it is unclear whether the group is moving to the West,
as suggested by the appearance of the X-ray residuals, or
to the East, as suggested by the positions of NGC4911 and NGC4921
(Neumann et al. 2003).  
Overall, this strongly suggests that the k+a spectra, i.e. the
truncation of the star formation activity in these galaxies and
possibly the previous starburst, could be the result of an interaction
with the hot intracluster medium (ICM).  

In contrast, the location of the red k+a galaxies in Fig.~1 does not appear 
to be correlated with the X-ray residuals.
The red k+a phase has a timescale that is comparable to the core crossing
time in a cluster like Coma, and any signature of the link between the
truncation of star formation and the location within a substructure is
thus erased in the older k+a's, while it is still detectable in the
youngest subsample of blue k+a's.

It is instructive to note that looking for a spatial segregation
in the location of galaxies on the sky would not allow to
establish a correlation between
the star formation history of the k+a galaxies and the substructure:
the link with the dynamical history of Coma
appears evident only once a
detailed X-ray map reveals the complicated structure in the hot
intracluster gas.

The blue k+a's do show, however, a radial velocity distribution that
is significantly different from that of the red k+a's and the global
Coma dwarf population.  Their mean radial velocity
is 8120$\pm 709$ km $\rm s^{-1}$, with all but one at $v>7200$ km $\rm
s^{-1}$.  In contrast, both the red k+a's and all faint galaxies with
passive spectra have much lower mean velocities: 6992$\pm 761$ and
6854$\pm 244$ km $\rm s^{-1}$, respectively.  


\section{Implications}

Two main conclusions can be drawn from the analysis of the Coma k+a galaxies. 

a) {\it The relation between k+a galaxies and substructure within the Coma 
cluster.}
The position of faint post-starburst galaxies in Coma relative
to X-ray substructure in the cluster strongly suggests that
the interruption of the star formation activity in these galaxies 
is a cluster-related phenomenon, most likely due to
the impact with the intracluster medium of recently infallen galaxy groups.
Moreover, the physical
process that caused the halting of the star formation must
have acted on a {\it short timescale}: such timescale must be significantly 
shorter than 1 Gyr, the timescale for the visibility of the k+a
signature in the spectra, for a k+a spectrum to be produced.

b) {\it Comparing the properties of k+a galaxies in Coma and
in clusters at z=0.4-0.5: the downsizing effect.}

Numerous spectroscopic surveys of galaxies in distant clusters have
found significant populations of luminous k+a galaxies (see for
example Couch \& Sharples 1987, Abraham et al. 1996, Dressler \& Gunn
1992, Fisher et al. 1998, Dressler et al. 1999, Tran et al. 2003, but
see also Balogh et al. 1999).  However, k+a galaxies as luminous as
those in distant clusters ($M_V \leq -20$) are absent in Coma,
where k+a's are detected at magnitudes typically fainter than $M_V \sim -18$
(P04). 

Post-starburst galaxies in Coma are thus much fainter than those
observed in distant clusters. K+a spectra in clusters appear
to be a luminous phenomenon at z=0.5 and a faint one at z=0.
In Coma we are observing late-type starforming galaxies becoming dwarf 
spheroidals, while the descendants of k+a's at high redshift
will be among the most massive early-type galaxies today (see also Tran et al.
2003, and these proceedings).
The observed evolution of the maximum luminosity of k+a galaxies
most probably reflects
a change in the galaxy populations infalling in clusters, and 
provides further evidence of a ``downsizing effect'': going to lower 
redshift, the maximum
luminosity/mass of galaxies with significant star formation activity 
progressively decreases, in {\it all environments}, and active
star formation in low mass galaxies seems to be more protracted on average
than in massive galaxies.

\section{Related (mostly open) questions}

1. Based on the strength of the lines, 
most of the blue k+a galaxies in Coma are inequivocally post-starburst
systems, in which a starburst preceded the halting of the star formation.
{\it Was the starburst triggered by the impact with the intracluster medium?}
Based on timescale arguments, it is possible that the interaction with the 
ICM produces both the starburst and the subsequent halting of the star 
formation. Low--mass star--forming galaxies, however, often have
a burst-like, non--regular star formation history, and thus the possibility 
that the starburst could be an intrinsic non--environmentally related 
phenomenon cannot be excluded.

\smallskip

2. {\it Could the termination of the star formation activity in these galaxies
be due to ``strangulation''?} 
The $\rm H\delta$ strength of the blue k+a
galaxies implies that star formation was truncated in these galaxies
on a short timescale, i.e. short compared to the k+a timescale of
1-1.5 Gyr. A slowly declining star formation activity such as
that envisaged if galaxies simply lost their gas halo reservoir when
becoming part of a group (``strangulation'', e.g. Bower \& Balogh
2003) is not able to produce such strong Balmer lines.

\smallskip

3. {\it Do k+a galaxies in distant clusters originate from the same physical 
process (the interaction with the ICM) that produced k+a galaxies
in Coma?} This is yet unknown, and needs to be uncovered by studies
of distant clusters (e.g. see Dressler et al. these proceedings).

\smallskip

4. {\it Are there k+a dwarf galaxies in distant clusters?}
So far, spectroscopy of distant cluster galaxies has only been
obtained for relatively bright galaxies, thus it is unknown
how many faint k+a's exist in clusters at high-z, and whether the 
luminous ones that have been observed 
are only the ``tip of the iceberg''.

\smallskip

5. {\it Are {\it field} k+a galaxies at low redshift dwarf galaxies too?}
Giant, massive k+a galaxies are known in the field in the local Universe
(e.g. Yang et al. 2004).
Most likely field k+a's originate from a different mechanism (such as
galaxy-galaxy mergers, Zabludoff et al. 1996) that has the same final 
effect of the dense environment: terminate the star formation activity.

\smallskip

6. In hierarchical models, small structures form first, and as a 
consequence low mass galaxies can be expected to have stellar populations
 on average {\it older} than more massive galaxies.  
{\it Is the result presented here
thus inconsistent with the predictions of semi-analytic
modeling within a hierarchical scenario?}
This depends on the prescriptions adopted for the star formation recipe.
Some recent models (e.g. De Lucia et al. in preparation) 
assume a star formation 
efficiency that increases with the halo mass, and thus obtain
a trend of increasing age with galaxy mass for
elliptical galaxies.

\smallskip

7. And finally, {\it what is the origin of the downsizing effect, i.e. of the
fact that star formation in low
mass galaxies appears to be more protracted on average than in massive
galaxies}? Most likely, since this phenomenon is known
both in clusters (e.g. Smail et al. 1998, Kodama \& Bower 2001,
De Lucia et al. 2004) 
and in the field (e.g. Cowie et al. 1996, Kauffmann
et al. 2003), the mass dependence of the star formation history 
we observe in clusters simply reflects the evolution of the 
galaxy populations infalling from the field and it
is intrinsic to the galaxy itself (e.g. related to its mass and due 
to the combination of gravity and self--regulating star formation, 
see Chiosi \& Carraro 2002), and not an environmental effect.


\end{document}